\documentclass[article,twocolumn]{revtex4-1}

\usepackage{graphicx}
\usepackage{dcolumn}
\usepackage{bm}

\usepackage[utf8]{inputenc}
\usepackage[T1]{fontenc}
\usepackage{mathptmx}
\usepackage{etoolbox}
\usepackage{soul}
\usepackage{graphicx}
\usepackage[normalem]{ulem}
\usepackage{natbib}
\usepackage{amsmath,array,amssymb}
\usepackage{dcolumn}
\usepackage{bm}
\usepackage{hyperref}
\usepackage{color}
\usepackage{amsmath, amsthm, amssymb, amsfonts}
\usepackage{todonotes}
\usepackage[]{mdframed}

\newcommand{\ket}[1]{|#1\rangle}
\newcommand{\bra}[1]{\langle #1|}
\newcommand{\beq}{\begin{equation}}
\newcommand{\eeq}{\end{equation}}
\newcommand{\bea}{\begin{eqnarray}}
\newcommand{\eea}{\end{eqnarray}}
\newcommand{\bma}{\begin{subequations}}
\newcommand{\ema}{\end{subequations}}
\newcommand{\ba}{\begin{array}}
\newcommand{\ea}{\end{array}}

\graphicspath{{img/}}

\newcommand*{\img}[1]{%
    \raisebox{-.1\baselineskip}{%
        \includegraphics[
        height=2.1\baselineskip,
        width=2.1\baselineskip,
        keepaspectratio,
        ]{#1}%
    }%
}

\begin{document}

\title[]{Quantum networks using rare-earth ions}

\author{Wolfgang Tittel}
\affiliation{Department of Applied Physics, University of Geneva, 1211 Geneva 4, Switzerland}
\affiliation{Constructor Institute of Technology,  8200 Schaffhausen, Switzerland}
\affiliation{Constructor University, 28759 Bremen, Germany}
\author{Mikael Afzelius}
\affiliation{Department of Applied Physics, University of Geneva, 1211 Geneva 4, Switzerland}


 \author{Adam Kinos}
\affiliation{Department of Physics, Lund University, P.O. Box 118, SE-22100 Lund, Sweden}

 \author{Lars Rippe}
\affiliation{Department of Physics, Lund University, P.O. Box 118, SE-22100 Lund, Sweden}

 \author{Andreas Walther}
\affiliation{Department of Physics, Lund University, P.O. Box 118, SE-22100 Lund, Sweden}

\date{\today} 

\begin{abstract}
We review concepts and recent work related to creating light-matter interfaces for future quantum networks based on rare-earth ion-doped crystals. More precisely, we explore their unique suitability for creating photon sources, optical quantum memories for light, and qubits that allow quantum information processing. In addition, we review the state-of-the-art of elementary quantum repeater links, and provide suggestions for future research.
\end{abstract}

\maketitle

\section{\label{sec:Intro}Introduction}

 The ever-increasing possibility to control nature at a scale where quantum effects arise creates opportunities that we could only dream of a few years ago. It has allowed understanding some of the most counter-intuitive predictions of quantum theory such as entanglement -- distant particles whose properties are tied together \cite{schrodinger1935,bell1964,aspect1981,tittel1998,weihs1998,rowe2001,hensen2015}. In turn, this has spurred a world-wide race to create the quantum internet \cite{wehner2018}. This revolutionary network promises provable-secure communication using quantum key distribution (QKD) \cite{gisin2002}, networked \cite{vanmeter2016} and blind quantum computing \cite{fitzsimons2017}, and distributed quantum sensing \cite{guo2020}. Common to all is the need for light-matter interfaces that employ atoms or optically addressable centers \cite{kimble2008,awschalom2018,reiserer2022}, ideally in the solid-state environment. For instance, single emitters allow the creation of single photons and  multi-photon entangled states \cite{tomm2021,senellart2017,zhang2008,cogan2023} as well as spin-photon and spin-spin entanglement \cite{barrett2005,hensen2015,knaut2024}; a collection of interacting centers enables quantum gates for quantum information processing \cite{bradley2019}; and large ensembles can be used for photon-multiplexed quantum memory \cite{sanders2009}, which is key to quantum repeaters and hence long-distance quantum communication \cite{briegel1998,sangouard2011}. For compatibility and scalability, these components should ideally be based on the same type of defect and be on-chip integrable using standard photonics technology. 

Due to their unique combination of suitable energy levels with long lifetimes, large inhomogeneous broadening, and remarkable optical and spin coherence times, ensembles of rare-earth ions doped into inorganic crystals have emerged during the past decade as excellent choices for optical quantum memories \cite{heshami2016}. In addition, their long spin coherence times---up to hours in the case of Eu:Y$_2$SiO$_5$ \cite{zhong2015}---make them highly suitable for encoding long-lived qubits, enabling the creation of spin-photon or spin-spin entangled states \cite{uysal2024,ruskuc2024}. Furthermore, the possibility for controlled dipole-dipole interactions underpins their potential for multi-qubit quantum information processing \cite{kinos2021}. However, the long excited-state lifetimes of rare-earth ions---a prerequisite for the long optical coherence time needed for multiplexed quantum memories---also represent a caveat as they hamper the observation of spontaneous emission from individual ions. This impacts the observation of single photons, and, by extension, the creation of entangled states and of optical readout of individual qubits. This problem can be solved by exploiting the Purcell effect, which allows increasing the ions’ emission rate by coupling it to a mode of a nano- or micro-cavity \cite{dibos2018,zhong2018,chen2020,ulanowski2022,gritsch2023,deshmukh2023,yu2023,huang2023,ourari2023,yang2023} (see ref. \onlinecite{kolesov2012} for an alternative approach). In turn, this opens the path toward the creation of compatible and on-chip integratable single-photon emitters,  individual ion quantum memory for light, and long-lived qubits based on the same material platform.

This review is organized as follows. First, we will give a general introduction to quantum networks and quantum repeater architectures whose developments are currently pursued by many groups worldwide. Next, we will briefly discuss the properties of rare-earth ions that underpin their potential as light-matter interfaces for future quantum networks. Sections \ref{sec:photons}-\ref{sec:nodes} describe three different use cases: single-photon sources, quantum memory for light, and single qubits that allow quantum information processing. In section \ref{sec:elementary-repeaters}, we will review recent work towards elementary quantum repeater links. Finally, we give an outlook with a brief discussion of remaining challenges and future research directions in section \ref{sec:outlook}.  

 \begin{figure*}[tt]
\includegraphics[width=0.9\textwidth]{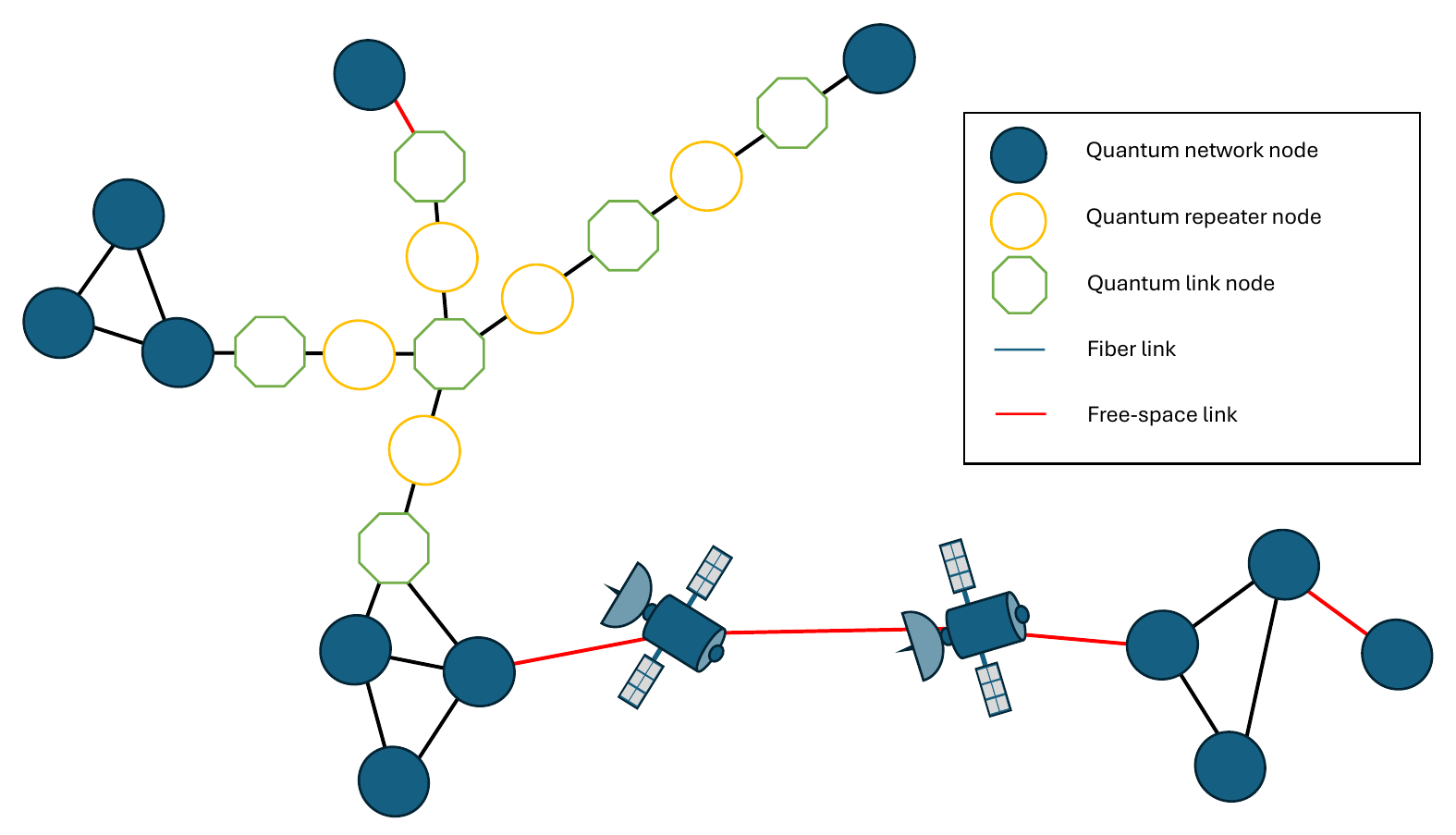}
\caption{\textbf{Quantum network.}  Illustration of a generic quantum network composed of network nodes, repeater nodes, link nodes, and fiber- as well as free-space (including satellite-based) photonic links.}
\label{fig-network}
\end{figure*}

\section{\label{sec:reps&networks}Quantum repeaters and quantum networks} 
\subsection{\label{subsec:networks}Quantum networks}
For this article we assume that a quantum network consists of \emph{network nodes} that allow the execution of quantum applications, and \emph{photonic communication links} that enable the exchange of quantum information across fiber and free-space (including satellite-based) quantum channels. To compensate for transmission loss, these links may use quantum repeaters and include \emph{quantum repeater nodes} and \emph{quantum link nodes}.

While some basic benefits can already be retrieved from quantum networks composed of network nodes that can only emit attenuated laser pulses and perform single qubit projection measurements---QKD over trusted nodes is a well-known example \cite{peev2009,dynes2019,chen2021,chen2021b,idQ}---here we assume fully quantum-enabled nodes that can emit quantum states of light, contain one or several long-lived qubits, and allow single and multi-qubit unitary operations (gates) and measurements. These nodes are connected through fiber or free-space channels into a small, e.g. metropolitan area quantum network, and allow executing general quantum network applications. Network nodes may also be connected to \emph{quantum link nodes} that allow quantum communication with other---distant---network nodes through chains of quantum \emph{repeater nodes} and additional \emph{quantum link nodes}. Finally, to span long distances---e.g. inter-continental distances---we imagine connections to and between satellites that may relay quantum information using direct optical links, repeater-based links, or even by physically transporting stored quantum information from one place to another. The cartoon in Fig. \ref{fig-network} shows a generic example of such a network. Note that some nodes may play more than one role, e.g. quantum networks nodes may double as quantum repeater nodes or link nodes.

\begin{figure*}
\includegraphics[width=1.5\columnwidth]{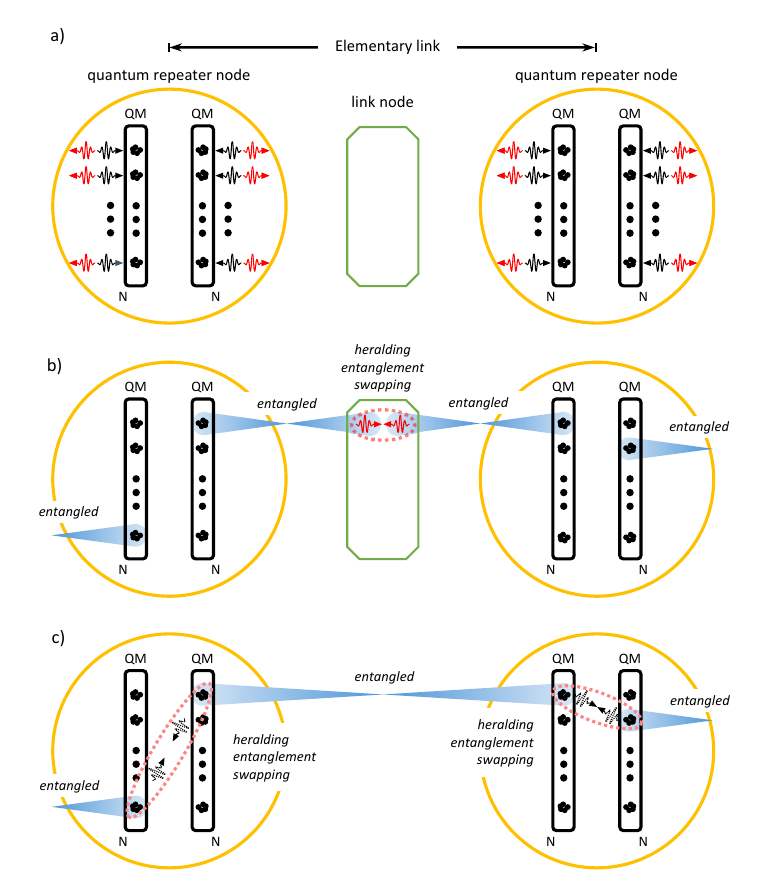}
\caption{\textbf{A quantum repeater based on atomic ensembles.} a) A long quantum communication link is split into shorter \emph{elementary links}. Each repeater node (yellow ellipse) emits 2N multiplexed entangled photon pairs (\img{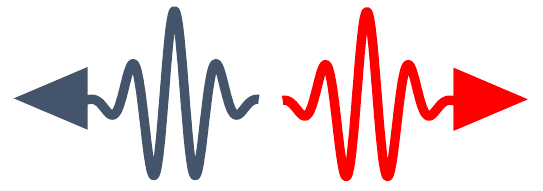}), and contains two multi-mode quantum memories (QM) that store one photon per emitted pair. Any degree of freedom is possible for multiplexing, e.g. time, frequency, polarization, and spatial modes. The 2nd photon per pair is sent to the link node. b) Photons meet at the link node and are pairwise subjected to a heralding entangling operation. Entanglement is indicated by the lying, blue, ‘figure of eight’, the entangling operation by a dashed red circle. Note that these steps are probabilistic : the travelling photons experience loss on their way from the repeater node to the link node and the entangling operation, a linear-optics Bell-state measurement, does not always project onto an entangled state \cite{lutkenhaus1999}. The figure assumes success in the 1st mode of the central elementary link, and in mode N and 2 in the left and right neighbouring link, respectively. c) The pairwise entanglement between photons and memories is ‘swapped’ to heralded entanglement between memories in adjacent nodes. The relevant atomic excitations are finally reconverted into photons and a mode mapping operation that makes photons from different elementary links indistinguishable is applied (not shown). Note that this part requires the memories to store photons during at least the time it takes the other member of the pair to reach the link node and classical information about the measurement result, and hence the information that determines the mode mapping operation to travel back. Finally, another entangling operation---again a probabilistic linear-optics-based Bell-state measurement---extends entanglement across adjacent links, again in heralded fashion.}
\label{fig-repeater}
\end{figure*}

\subsection{\label{subsec:repeaters}Quantum repeaters}
Entanglement is the physical property that marks the most striking deviation of the quantum from the classical world: highly---if not perfectly---correlated measurement results of particles that can in principle be arbitrarily far away. As such, entanglement is a fundamental resource empowering future quantum networks, allowing, e.g., provable-secure establishment of cryptographic keys by means of quantum key distribution (QKD) \cite{gisin2002} and the faithful transfer of an unknown quantum state between distant nodes using quantum teleportation \cite{bennett1993}. Unfortunately, photons---the particle of choice to distribute entanglement---are subject to loss and decoherence as they travel through optical fibers, limiting the extension of quantum networks of the type described above to around 100 km. Solutions to this problem are quantum repeaters \cite{briegel1998,sangouard2009}, the transmission of entangled photons to distant locations using satellites instead of fibers \cite{yin2017a}, or a combination thereof. 

\begin{figure*}
\includegraphics[width=1.5\columnwidth]{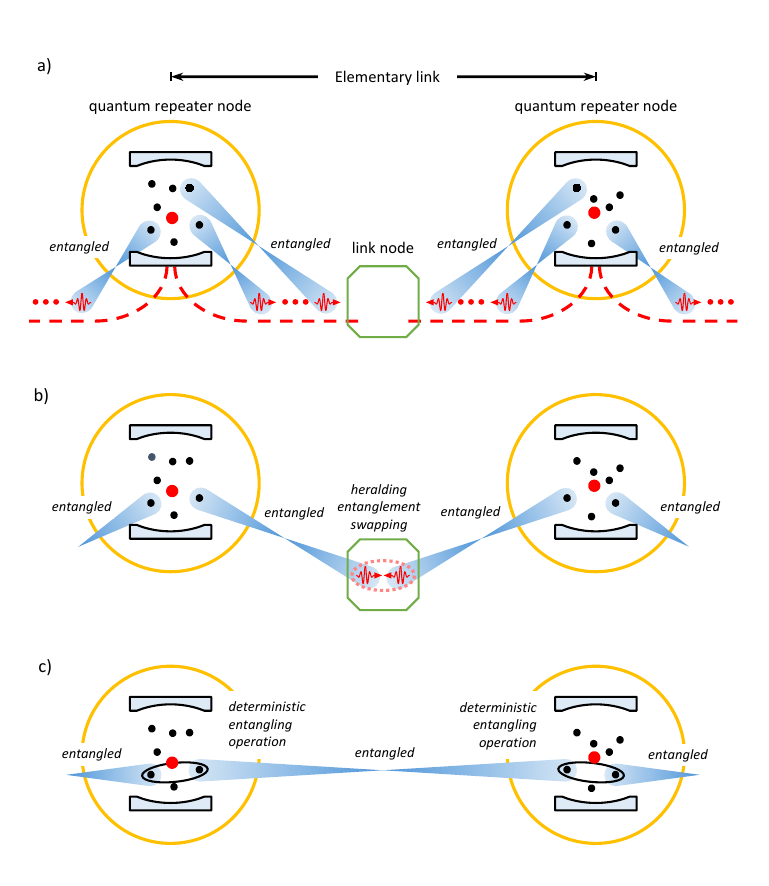} 
\caption{\textbf{A quantum repeater based on individual ions.}  a) As in Fig. \ref{fig-repeater}, a long quantum communication distance is split into shorter \emph{elementary links}. However, here each repeater node contains a cavity with individual ions that encode qubits, instead of atomic ensembles. Each ion is entangled with a single photon (depicted by the lying, blue, ‘figure of eight’), which is sent to a neighbouring link node. b) At the link node, photons are pairwise subjected to a heralding entanglement swapping operation (dashed red circle). This step is  probabilistic, with the figure showing one of the successful entangling operations. c) The entanglement between individual ions and photons is now extended to two individual ions in adjacent repeater nodes. Finally, by using two-qubit operations and measurements on two individual ions inside a single node, deterministic entanglement swapping (solid black circles) is performed to extend the entanglement across the quantum communication channel. This deterministic entangling operation and the ability to perform other qubit operations e.g. entanglement purification are  key advantages of this architecture compared to the approach based on atomic ensembles in Fig. \ref{fig-repeater}. In all parts the red ion indicates a dedicated communication ion used in some of the protocols, see section \ref{sec:nodes}.}
\label{fig-repeater2}
\end{figure*}

Figs. \ref{fig-repeater} and \ref{fig-repeater2} illustrate two different approaches to quantum repeaters. The first takes advantage of entanglement creation across \emph{elementary links}---the basic building block of a repeater-based quantum communication link---using highly multiplexed (ensemble-based) quantum memories and equivalently multiplexed photon-pair sources \cite{sinclair2014}, making them nearly deterministic. It furthermore employs entanglement connection (entanglement swapping) between neighbouring elementary links using linear-optics Bell state-measurements \cite{michler1996}. This measurement is generally assumed to be probabilistic\cite{lutkenhaus1999}, but note that it can in principle be achieved deterministically by adding auxiliary entangled photons \cite{grice2011,barz2023}. The second approach is based on the probabilistic creation of entanglement across elementary links using individual ions as qubits and single photons, and Bell-state measurements to connect adjacent links based on the deterministic interaction between two neighbouring rare-earth ions \cite{kinos2023a}. Using many qubits, this architecture also offers the possibility for multiplexing, however, at a smaller degree than the first approach. Note that it is in principle possible to combine the best of both worlds \cite{gu2024}, however, here we focus on architectures that are already under experimental development using rare-earth ions. But regardless of the approach, all these approaches derive their improved scaling compared to entanglement distribution without quantum repeaters from the possibility of connecting elementary links \emph{after} confirmation of entanglement distribution.

\begin{figure*}
\centering
\includegraphics[width=0.9\textwidth]{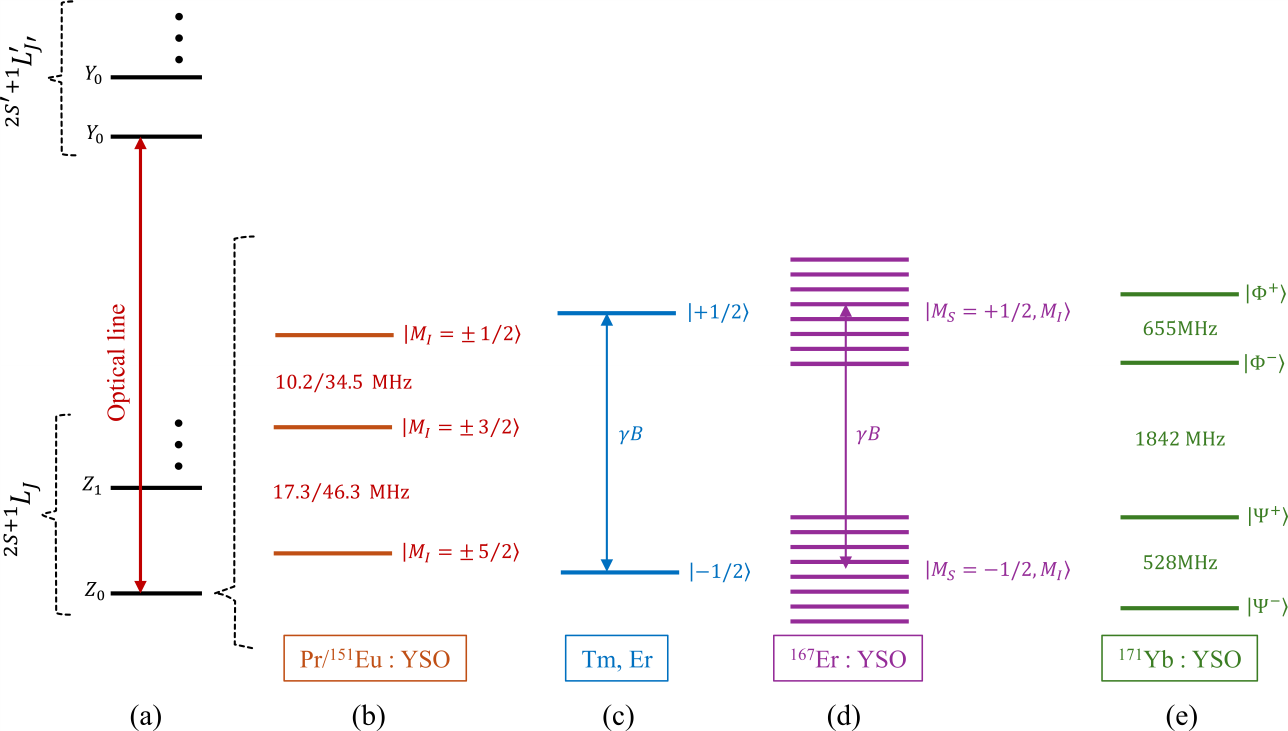}
\caption{\textbf{Rare-earth energy diagrams.} (a) The electronic states $^{2S+1}L_J$ and $^{2S'+1}{L'}_{J'}$ split into crystal field levels $Z_i$ and $Y_j$, each having electronic spin $S = 0$ for non-Kramers ions and $S=1/2$ for Kramers ions. The main optical transition of interest---the so-called zero-phonon line---connects $Z_0$ to $Y_0$. Panels (b)-(e) present some examples of Zeeman/hyperfine $Z_0$ manifolds (analogous for $Y_0$). (b) $S=0$ ions Pr$^{3+}$ and Eu$^{3+}$  with $I= 5/2$ feature three pseudo-quadrupole states in Y$_2$SiO$_5$ (YSO), each being a nuclear Zeeman doublet \cite{equall1995,yano1992}. (c) The $S=0$ ion Tm$^{3+}$ with $I = 1/2$ has a nuclear Zeeman doublet $\ket{M_I = \pm 1/2}$ with a crystal dependent $\gamma$ of the order of 1 GHz/T\cite{thiel2010}. $S=1/2$ ions with $I=0$ have electronic Zeeman doublets $\ket{M_S = \pm 1/2}$ with a large magnetic ratio $\gamma = \mu_b g_{eff}$, where $g_{eff}$ can reach $15$ for Er$^{3+}$ in Y$_2$SiO$_5$ \cite{sun2008} ($\mu_b \approx 14$ $\mathrm{GHz/Tesla}$). (d) The electronic hyperfine structure of $^{167}$Er$^{3+}$ in Y$_2$SiO$_5$, with $S=1/2$ and $I = 7/2$, at high magnetic field. It features a rich system with 8 hyperfine states per crystal field level \cite{rancic2017}. (e) For the $S=1/2$ and $I=1/2$ ion $^{171}$Yb$^{3+}$ in Y$_2$SiO$_5$, the $Z_0$ level splits into four non-degenerate Bell states, i.e. entangled states in the $\ket{M_S = \pm 1/2}$ and $\ket{M_I = \pm 1/2}$ spin projections, which are magnetic clock transitions\cite{ortu2018}. Note that examples (b) and (e) are given at zero magnetic field.}

\label{fig-levels}
\end{figure*}

\section{\label{sec:REIcrystals}Rare-earth ion-doped crystals}

The rare-earth elements are characterized by partially filled 4f shells. When doped into inorganic crystals, they generally form trivalent (triply positively charged) ions with inhomogeneously broadened 4f-4f transitions in the visible and near-infrared, with excited state lifetimes ($T_1$) that often exceed 1 ms. The crystal field interaction lifts the degeneracy of the electronic states given by $^{2S+1}L_J$ (where $2S+1$, $L$, $J$ are, in this order, the spin multiplicity, the angular momentum and the total angular momentum), see Fig. \ref{fig-levels}(a). Depending on whether the ion has an even (non-Kramers ion) or odd (Kramers ion) number of electrons, this results in $2J+1$ or $(2J+1)/2$ crystal field levels (or Stark levels) with effective electron spin $S=0$ and $S = 1$, respectively. The optical transition of interest typically couples the lowest crystal field levels of each electronic state, $Z_0 \rightarrow Y_0$, a so-called zero-phonon line (ZPL), as shown in Fig. \ref{fig-levels}(a). The $Z_0$ ($Y_0$) level has a Zeeman/hyperfine structure that depends on the electron spin $S$ and the nuclear spin $I$, resulting in a rich diversity of possible hyperfine manifolds according to the rare-earth element, its specific isotope, and the point symmetry of the doping site in the crystal. In Fig. \ref{fig-levels}(b)-(e), we give an illustrative list of examples, although not exhaustive, of possible manifolds.

At cryogenic temperatures $<$ 4 K, optical coherence times ($T_2$) of the optical ZPLs can approach $T_1$, and for some cases they exceed 1 ms \cite{equall1994,bottger2009,askarani2021}. This feature distinguishes rare-earth ions from all other solid-state emitters. Similarly, population lifetimes $T_1$ and coherence times $T_2$ between hyperfine levels within$Z_0$ manifold can be very long, with $T_1$ values of weeks \cite{konz2003,Oswald2018} and $T_2$ values of hours \cite{zhong2015} having been reported, respectively. For an introduction to the properties of rare-earth ion-doped crystals at cryogenic temperatures, see \cite{tittel2010,thiel2011,goldner2015,jacquier2005}.

\section{\label{sec:photons}Network components : Single and entangled photons}
\subsection{\label{subsec:SPDC}Spontaneous parametric downconversion} 

The current push by many countries to establish extended quantum networks is based on thorough understanding of its constituents and their interplay, as well as on a large number of proof-of-principle demonstrations of ever-increasing complexity by us and others. At the heart of this development have been sources of photon pairs based on spontaneous parametric down-conversion (SPDC) -- a process in which a strong pulse of light---the pump---is probabilistically converted inside a non-linear crystal into a pair of entangled photons, first demonstrated in 1970 \cite{burnham1970}, SPDC sources have rapidly gained a lot of attention both for fundamental tests of nature \cite{ou1988,tittel1998,weihs1998,giustina2015,shalm2015} as well as for applications such as quantum key distribution \cite{jennewein2000,naik2000,tittel2000} and quantum teleportation \cite{bouwmeester1997,boschi1998}, including deployed over optical fibers \cite{valivarthi2016,sun2016}, over free-space links between optical ground stations \cite{ma2012}, and even to a satellite \cite{ren2017}.

However, the state created by SPDC sources is not a true two-photon state, but rather a two-mode squeezed state, characterized by an even number of emitted photons. In qubit-based quantum communications one generally only considers the case with two photons (one pair), which describes the desired state $\ket{\psi}=(\ket{01}-\ket{10}/\sqrt{2}$. But the undesired higher-order contributions also exist, and it is unavoidable that 4 or 6 photons will sometimes be emitted. This problem is often countered by reducing the pump power so that the probability of creating more than 2 photons per pulse is small compared to creating one pair. However, this workaround faces important limitations in the case of a repeater. Due to the simultaneously increasing probability to create no photons at all, it is inefficient in the case in which the Bell-state measurements in Fig. \ref{fig-repeater} are based on the detection of a single photon. Worse, in case the projection onto a Bell state is indicated by the detection of two photons, it is not suitable at all \cite{guha2015}. Indeed, the best connection between two distant QKD nodes then uses only a single elementary link – concatenation of several links is not useful, and the repeaterless bound \cite{takeoka2014,pirandola2017}---a fundamental bound that describes the scaling of the secret key rate as a function of loss assuming no quantum repeater---cannot be violated. 

It is therefore important to develop better sources of entangled photons, a task that is further complicated by the need for spectral compatibility with optical quantum memories and telecommunication fibers. Several approaches have been proposed and/or implemented, including post selection of desired SPDC emissions \cite{reiserer2013,sinclair2016,krovi2016}, suppression of undesired emissions \cite{huang2012}; the use of individual emitters \cite{franson1989,huber2018}; and fusion of four (non-entangled) single photons---also created using individual emitters---into one heralded pair \cite{browne2005,zhang2008}. 
Motivated by the latter, we will focus in the following on the creation of single photons.

\subsection{\label{sec:Purcell}Cavity-enhanced single photon emission}

True single photons are highly valuable resources for quantum communication protocols such as quantum key distribution \cite{gisin2002}, but also for different quantum repeater architectures \cite{sangouard2011}, including as a resource for heralded entangled photon pairs \cite{browne2005,zhang2008}. They have been created using several types of solid-state emitters such as quantum dots \cite{tomm2021,senellart2017}, diamond vacancy centers \cite{bogdanov2018,ruf2021} and since 2018 also individual rare-earth ions. The latter are of particular interest as they constitute the only solid-state defect with long optical coherence time, a requirement for the creation of efficiently multiplexed quantum memory for light \cite{askarani2021}. 

\begin{figure}
\includegraphics[width=0.8\columnwidth]{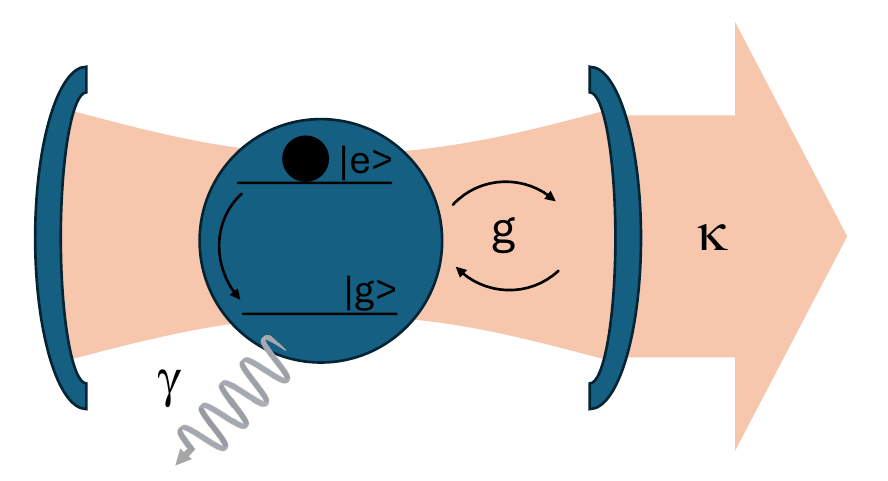}
\caption{\textbf{Cavity-enhanced single-photon source.}  Illustration of a single-photon source based on cavity-enhanced light-matter interaction. The cavity will typically be operated in the \emph{fast (or weak) cavity} regime. Here, the cavity decay rate $\kappa$ is large compared to the light-matter coupling constant $g$, which in turn is large compared to the spontaneous emission rate $\gamma$, i.e. $\kappa\gg g\gg \gamma$.}
\label{fig-Purcell}
\end{figure}

To create true single photons, the conceptually simplest approach is to excite a single atom, ion or optical centre, and to wait for subsequent spontaneous emission. However, due to their long excited state lifetimes---often ms---along with the fact that spontaneous emission is undirected, this approach is very inefficient for rare-earth ions. Yet, by coupling the ion to the mode of a cavity with small mode volume $V$ and large quality factor $Q$---resulting in the modification of the ion's electromagnetic environment---it is possible to increase the emission rate and furthermore to direct the emission into a mode that is defined by the cavity (see Fig. \ref{fig-Purcell}). This is called the \emph{Purcell effect} \cite{pelton2015}. The enhancement of the emission rate is given by the ratio of the cavity density of states to that of free-space---the Purcell factor $F_P$---and can be expressed by

\beq
F_P=\frac{3}{4\pi^2 }\beta \left (\frac{\lambda}{n}\right )^3  \frac{Q}{V}  \frac{|E(\vec{r} )|^2}{|E_{max} ^2|}.	
\eeq
Here, $\beta$ is the branching ratio of the desired transition,  $n$ is the refractive index of the crystal, and $E(\vec{r})$ and $E_{max}$ denote the field at the position of the ion and the maximum electric field, respectively. Note that we assumed optimized polarization. The reduction of the lifetime $T_1$ furthermore yields the potential of achieving radiation-limited emission with $T_2=2T_1$ (with coherence time $T_2$).

While difficult with traditional cavities, Purcell-enhanced emission and the observation of single photons from individual rare-earth ions has become possible due to the emergence of micro- and nano-scale cavities with mode volumes in the order of $\lambda^3$ or smaller. Starting with standard semiconductor platforms such as Si \cite{khriachtchev2016}, fabrication methods for photonic crystal nanocavities \cite{istrate2006} have lately been extended, e.g. to diamond \cite{burek2014,khanaliloo2015}, yttrium orthovanadate \cite{zhong2016} and LiNbO$_3$ \cite{rusing2019,li2019,boes2018,li2023}. In parallel, open microcavities with sufficiently small mode volumes have been developed \cite{hunger2010, deshmukh2023}, which would allow the use of any rare-earth material. 

\begin{figure*}
\includegraphics[width=0.8\textwidth]{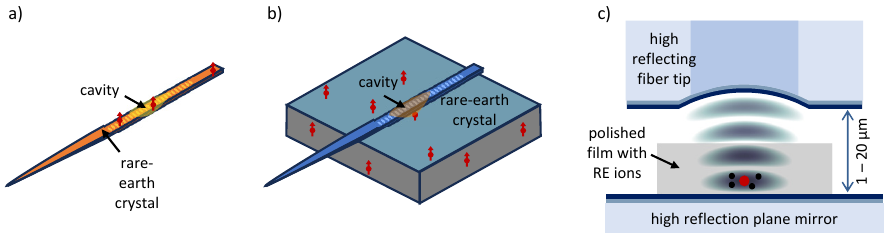}
\caption{\textbf{Purcell-enhanced single photon sources.} Purcell-enhanced emission into a nanocavity using the (a) homogeneous approach (in which the cavity is made out of the rare-earth crystal), (b) heterogeneous approach (in which the cavity field couples evanescently to rare-earth ions in a neighbouring crystal and (c) open microcavity approach (in which a rare-earth crystal is placed within an open microcavity).}
\label{fig-coupling}
\end{figure*}

To achieve Purcell-enhanced emission based on photonic crystal cavities, two approaches are being pursued (see Fig. \ref{fig-coupling}). First, it is possible to create the nanocavity directly out of the rare-earth crystal, either by means of focused ion beam milling \cite{zhong2018} or through reactive ion etching \cite{gritsch2023}. In this case, the coupling between the light and the rare-earth ion happens inside the cavity where the electric field is highest and $E(\vec{r})$ approaches $E_{max}$. Alternatively, one can also fabricate a cavity out of a material without rare-earth doping, and then transfer the cavity onto the rare-earth-doped crystal \cite{dibos2018}. In this “heterogeneous” approach, the coupling with a single ion is mediated via the evanescent field, which is smaller than in the “homogeneous” case where the ion is located within the cavity. Note that the addressable rare-earth-ion transitions are limited in both cases to the transparency window of the (external) cavity. In the case of silicon, currently the most widely used material, this excludes wavelengths below 1.1 $\mu m$ and hence the transitions in Eu ($\lambda\approx$ 580 nm, depending on the host crystal), Pr ($\lambda \approx$606 nm), Tm ($\lambda \approx$794 nm) and Yb ($\lambda\approx$980 nm) that are currently being investigated for optical quantum memory \cite{ortu2021,duranti2024,askarani2021,businger2020}. 

\subsection{\label{subsec:sources_figs_merit}Figures of merit}

\begin{figure}[bbbb]
\includegraphics[width=0.8\columnwidth]{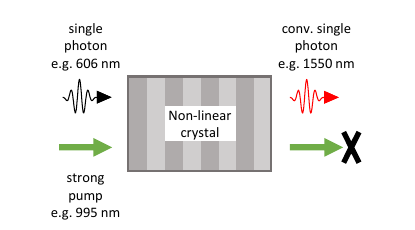}
\caption{\textbf{Quantum frequency conversion.} The interaction of a single photon with a strong laser (pump) beam within a non-linear crystal can lead to the conversion of the frequency of the photon by means of so-called three-wave or four-wave mixing without affecting the encoded quantum information.}
\label{fig-v_conversion}
\end{figure}

Figures of merit for single-photon sources include \emph{wavelength}, \emph{spectral bandwidth} and the possibility for \emph{multiplexing}, all of which determine compatibility with quantum memory and transmission over optical fiber or free-space links. Note that it may be necessary to change the wavelength of the emitted photons as it will in most cases only be compatible with either the absorption line of a quantum memory or the transparency window of the transmission medium, but not with both. An exception are erbium-based sources, which allow emitting photons at telecom wavelength of 1532 nm that are obviously also spectrally compatible with erbium-based quantum memories (see section \ref{subsec:QM_SOTA} for more info about Er-based memories). Otherwise, the required frequency conversion can be implemented by means of a non-linear interaction between the photon to be converted and a strong laser pulse \cite{tanzilli2005,dreau2018,maring2018}, as depicted in fig. \ref{fig-v_conversion}. Additional figures of merit include a high \emph{probability of creating true single photons}, the latter being generally assessed in terms of the autocorrelation coefficient $g^{(2)}(0)$, where, ideally, $g^{(2)}(0)$=0. 

Currently, the largest challenge is to achieve \emph{indistinguishability}, which is at the heart of multi-photon interference \cite{michler1996}. In turn, indistinguishability enables the creation of heralded entangled photons using four single photons \cite{zhang2008}, the entangling operation in the link nodes (see Figs. \ref{fig-repeater} and \ref{fig-repeater2}), and quantum teleportation between repeater nodes \cite{valivarthi2016,ruskuc2024}, to name just a few applications. Of particular concern is spectral purity, i.e. the realization of a Fourier-limited and stable spectrum, free of spectral diffusion.     

And finally, being an important criterion for any quantum technology to become a real-world application, single photon sources must become \emph{scalable and easy to use}.  

\begin{figure*}
\includegraphics[width=\textwidth]{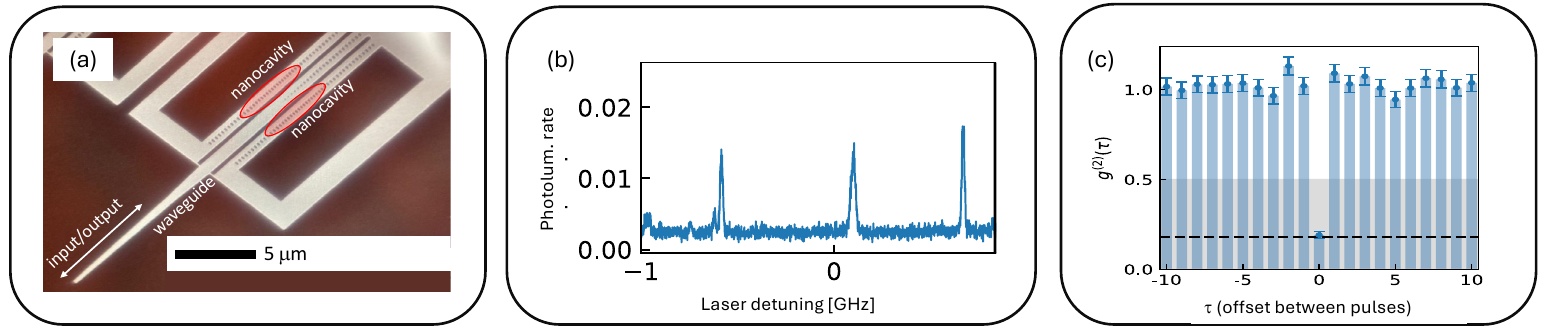}
\caption{\textbf{Single photon emitters.} Some examples of single photon emitters developed by the authors. (a)-(c) Scanning electron microscope picture of a silicon nano-photonic cavity coupled to an Er-doped LiNbO$_3$ crystal, isolated photo-luminescence lines, and result of autocorrelation measurement (Tittel/Gr\"oblacher group \cite{yu2023}).}
\label{fig-SEM}
\end{figure*}
    
\subsection{\label{subsec:sources_SOTA}Single photon sources -- state-of-the art} 
  
Starting with the first two demonstrations in 2018 \cite{zhong2018,dibos2018}  single-photon detection based on Purcell-enhanced light-matter interaction has now been reported by several groups \cite{chen2020,merkel2020,ulanowski2022,xia2022,gritsch2023,deshmukh2023,yu2023,huang2023,ourari2023,yang2023,ruskuc2024} (see fig. \ref{fig-SEM} for an example and ref. \onlinecite{kolesov2012} for an alternative approach). However, only two groups could so far demonstrate the indistinguishability of subsequently emitted photons from the same rare-earth ion \cite{ourari2023, ruskuc2024}. Furthermore, the possibility of quantum interference with photons from two different sources, which underpins the functioning of quantum networks and especially of quantum repeaters, has only been reported once \cite{ruskuc2024}. The problem is spectral diffusion, which leads to broadening of the photons’ spectra beyond the limit imposed by the radiative emission rate $\gamma$. This results in distinguishability and prevents two-photon interference. 

Spectral diffusion is a common problem for all solid-state emitters and has been suggested to be particularly pronounced for rare-earth ions close to surfaces \cite{ourari2023}, as is the case in nanophotonic structures. To limit spectral diffusion due to time-varying electric fields, the above demonstrations of indistinguishability and interference \cite{ourari2023, ruskuc2024} were therefore based on crystals in which the rare-earth ions occupy sites with non-polar symmetry. Another---or an additional---strategy is to use open micro-cavities as in fig. \ref{fig-coupling}(c), where relatively thick crystals allows interacting with rare-earth ions that are far away from surfaces \cite{ulanowski2022}. However, some authors reported that no such ``nanostructure-enhanced'' spectral diffusion has been observed in crystals with polar symmetry where the effect should be more pronounced \cite{dutta2023}. In addition, it has also been proposed that the use of ions within the tail---not the center---of the inhomogeneously broadened absorption line, for which strain may impact the site symmetry, could lead to additional spectral sensitivity \cite{kindem2020}. 
This leaves the question of how to minimize spectral diffusion in order to create indistinguishable single photons currently open.


\section{\label{sec:Memory}Network components : Ensemble-based quantum memory for light}
The quantum repeater in Fig. \ref{fig-repeater} requires one to delay---or store---photons until the results of measurements of other photons become available. The minimum required delay is given by the round-trip communication time set by the speed of light, but can be significantly longer due to photon transmission loss and other inefficiencies \cite{simon2007,sangouard2011}. The storage of photons allows conditional operations, also referred-to as feed-forward control. To optimize the entanglement distribution rate, it should be possible to continuously add photons to the memory, i.e. the memories should be multimode \cite{simon2007,sinclair2014}. Atomic ensembles with a large number of atoms such as rare-earth ion-doped crystals are natural candidates for the required storage as they allow one to reversibly map a large number of temporally, spectrally or spatially multiplexed photons onto different collective atomic modes with negligible overlap \cite{nunn2008,afzelius2009,ortu2022}.


\subsection{\label{subsec:PEQM}Photon echo quantum memory with rare-earth-ion ensembles}

An important class of quantum memory (QM) protocols that lends itself ideally to materials with inhomogeneously broadened optical absorption lines such as rare-earth doped crystals stems from the photon-echo protocol \cite{abella1966}. ‘Photon-echo quantum memory’ relies on the transfer of the optical quantum state onto a collective atomic excitation. Assuming the absorption of a photon in an ensemble of two-level atoms at time $t_0$, the atomic state becomes \cite{lvovsky2009}
\beq
\ket{\psi}_A=\frac{1}{\sqrt{N}}\sum_{j=1}^N c_j e^{i2\pi\delta_j (t-t_0)} e^{-ikz_j}\ket{g_1, ..., e_j, ..., g_N}	 
\label{eq:Dickestate}
\eeq
where $N$ is the total number of atoms, and $g_j$ and $e_j$  denote the ground and excited state, respectively, of atom $j$. The wave number of the optical field is given by $k$, $z_j$ is the position of atom $j$, and $c_j$ characterizes  the frequency and position-dependent coupling of atom $j$ to the field. Since we are dealing with an inhomogeneous ensemble, each atom has a different detuning $\delta_j$ of the atomic transition with respect to the optical carrier frequency, which causes inhomogeneous dephasing of the state in Eq. \ref{eq:Dickestate} after the absorption of the photon. Often seen as a nuisance, this can in fact be used as a resource for multiplexing. For instance, in the time domain, $l$ photons absorbed at different times $t_0^l$ will each create collective states of the same form as in Eq. \ref{eq:Dickestate}, but they are distinguishable by their different start times $t_0^l$. Similarly, photons in different spatial modes $k$ will generate distinguishable collective states. To map these atomic excitations back onto optical modes, i.e. to trigger the re-emission of the stored photons, the inhomogeneous dephasing must be undone. All photon-echo type QM protocols employ a specific method to achieve the required rephasing. In the following we will focus on the \emph{atomic frequency comb} (AFC) scheme ----the currently most widely implemented protocol--- but we will also briefly mention alternative protocols.

\begin{figure}[]
\includegraphics[width=\columnwidth]{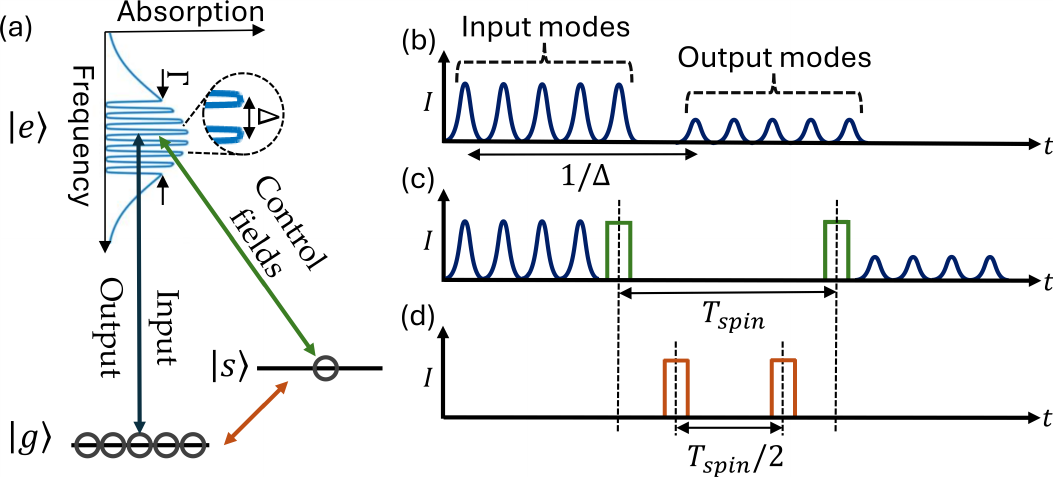}
\caption{\textbf{The AFC protocol.} (a) The energy level diagram of the AFC quantum memory. The scheme involves a preparation step where ions are first polarized into the $\ket{g}$ state, and the comb structure with a bandwidth $\Gamma$ is created on the optical $\ket{g} - \ket{e}$ transition by spectral hole burning methods. (b) The 2-level AFC scheme stores the input modes as optical excitations and the output modes are emitted after a pre-determined storage time $1/\Delta$. (c) The 3-level (or spin-wave) AFC allows storing the input modes for longer durations using a hyperfine level $\ket{s}$, and on-demand read-out by the control fields (eg. $\pi$ pulses). (d) The spin storage time $T_{spin}$ can be extended further by applying spin $\pi$ pulses (spin control).}
\label{fig-AFC}
\end{figure}

The AFC protocol is based on an inhomogeneously broadened absorption line tailored into a periodic series of narrow absorption peaks \cite{afzelius2009} with detuning $\delta_j = m_j \Delta$, where $m_j$ is an integer and $\Delta$ is the comb periodicity, see Fig. \ref{fig-AFC}(a). Such an atomic frequency comb can be created by persistent spectral hole burning (SHB) techniques, where recently developed techniques have allowed the creation of broadband and high resolution AFC structures \cite{jobez2016,businger2022}. An incoming photon is absorbed by the comb, with the comb bandwidth $\Gamma$ matching the photon bandwidth. Thanks to the comb periodicity, it follows directly from Eq. \ref{eq:Dickestate} that the collective state rephases at a time $t = 1 / \Delta$, causing photon re-emission (output) in the form of the AFC echo, see Fig. \ref{fig-AFC}(b). This 2-level AFC echo acts like a delay line with pre-determined storage time, the on-demand read out based on the 3-level AFC will be discussed below.

A key feature of the AFC scheme is its high temporal multimode capacity. The number of temporal modes $N_t$ that can be efficiently stored only depends on the number of teeth in the comb $N_{teeth}$, the capacity being $N_t = N_{teeth}/2.5$ \cite{ortu2022}. To maximize the temporal mode capacity, one should thus look for materials with narrow homogenous linewidth, i.e. long optical coherence time, and large absorption bandwidth. 

The efficiency of an ensemble-based quantum memory depends on the collective light-matter coupling, i.e. it depends on the optical depth of the ensemble \cite{sangouard2007,gorshkov2007,afzelius2009}. Here we need to distinguish forward and backward read-out of the re-emitted photon with respect to the input photon. For forward recall, the efficiency for the echo at $t = 1 / \Delta$ is $\eta_{AFC} = \tilde{d}^2 \exp(-\tilde{d}) \eta_d$ \cite{afzelius2009}, where $\tilde{d}$ is the effective optical depth averaged over the AFC and $\eta_d$ a dephasing factor determined by the shape of a single tooth \cite{afzelius2009}, whose width is limited by the inverse coherence -- the homogeneous linewidth. For backward recall, which can be achieved by phase matching and auxiliary control fields \cite{tittel2010}, the efficiency is given by $\eta_{AFC} = (1- \exp(-\tilde{d}))^2 \eta_d$. For forward recall, the efficiency is limited to 54\%  due to the re-absorption factor $\exp(-\tilde{d})$ \cite{sangouard2007}, while for backward recall it can approach 100\% for high optical depth and high comb finesse \cite{sangouard2007,afzelius2009}. An alternative method to reach high efficiency is to put the memory in a single-sided optical cavity whose input mirror is impedance matched to the memory absorption \cite{afzelius2010}. In this way, unit efficiency can be approached without resorting to the phase matching operation of backward recall, even for a moderate cavity finesse. For a perfectly impedance-matched, loss-less cavity, the efficiency is only limited by the intrinsic AFC dephasing factor $\eta_{AFC} = \eta_d$ \cite{jobez2014}, which can approach unity for high enough AFC finesse.

Let us point out that an alternative rephasing method, the so-called gradient echo memory (GEM) protocol \cite{hetet2008} (which is closely related to the \emph{controlled reversible inhomogeneous broadening} (CRIB) protocol \cite{kraus2006,alexander2006}), allows forward recall that is not limited by re-absorption. However, GEM/CRIB memories have less temporal multimode capacity with respect to AFC memories \cite{nunn2008}.

The collective state in eq. \ref{eq:Dickestate} can also be rephased using optical $\pi$ pulses. The method called \emph{revival of silenced echo} (ROSE) \cite{damon2011} is a variant of the two-pulse photon echo and avoids in principle population inversion and hence spontaneous emission noise during the echo emission. However, in practice, imperfect $\pi$ pulses cause such noise \cite{bonarota2014}, and the potential of ROSE for quantum state storage remains an open question. To overcome this problem, the \emph{noiseless photon echo} (NLPE) protocol was recently proposed \cite{ma2021b}, where a 4-level system is employed instead of the 2-level system in ROSE. The advantage of NLPE with respect to AFC is that no initial memory preparation step (spectral hole burning) is required, which removed complexity and leads to higher efficiencies thanks to a higher effective optical depth \cite{ma2021b}. However, the 4-level scheme adds dephasing with respect to the AFC scheme due to uncorrelated inhomogeneous broadening of the employed transitions, which reduces the optical storage time and hence the multimode capacity with respect to an AFC memory in the same material \cite{ma2021b}.

For a quantum memory that exploits a single optical transition, the storage time is limited by the optical coherence time T$_2$. In the case of rare-earth crystals, T$_2$ can reach ms \cite{koenz2003,bottger2009,askarani2021,nicolas2023} -- a unique feature among solid-state centers. Even longer storage times can be achieved when mapping the optical coherence in a reversible manner onto hyperfine states using optical control $\pi$-pulses -- a technique that is generally referred-to as 3-level (or spin-wave) AFC protocol \cite{afzelius2009,afzelius2010b}, see Fig. \ref{fig-AFC}(c). Moreover, for AFC memories this also allows readout-on-demand, which is not possible in case of the conventional 2-level AFC scheme, although some degree of control of the read-out time can be achieved with an additional Stark effect control \cite{horvath2021, liu2020}. A challenge for such spin-wave mapping is to achieve efficient control fields over the memory bandwidth given the weak optical transition dipoles in rare-earth ions. This problem can be solved using frequency-chirped adiabatic pulses \cite{minar2010,chen2024}. In the simplest version, the spin-wave memory storage time is limited by inhomogeneous dephasing of the hyperfine transition, typically on the order of tens of $\mu$s \cite{afzelius2010b,gundogan2015}. The storage time can be extended to the spin coherence time by applying two $\pi$ pulses on the hyperfine transition \cite{jobez2015}, see Fig. \ref{fig-AFC}(d). By dynamically decoupling the hyperfine transition from the perturbing environment using many $\pi$ pulses the storage time can be extended further \cite{holzaepfel2020}, possibly up to the ultimate limit of $2 T_1$.

To finish this section, let us also mention ``photon emission"-based approaches for multimode light-matter entanglement generation, i.e. the RASE \cite{ledingham2010} and AFC-DLCZ protocols \cite{sekatski2011}. Here, the atomic system acts as both a source and memory for pairs of photons. These schemes have large temporal multimode capacity and include the possibility for extended spin-wave storage. However, the sources are probabilistic, featuring the same issues as sources based on SPDC (see sec. \ref{subsec:SPDC}). Furthermore, storage via electromagnetically induced transparency (EIT) has been explored for long-duration single photon storage \cite{hain2022}. Yet, due to their limited  temporal multimode capacity \cite{nunn2008}, EIT-based memories are less attractive for quantum networks. 

\subsection{\label{subsec:QM_fig_merit}Figures of merit}
A large number of figures of merit can be used to characterize the suitability of a quantum memory for a quantum repeater link. Below we discuss the most frequently used ones, and provide some information relevant to their realization using the AFC quantum memory protocol in rare-earth-ion doped crystals.

\emph{Storage efficiency}. The memory efficiency is a key metric for quantum repeaters, as the final rate will scale with the memory efficiency to the power of the number of memories employed in the repeater link (the exact scaling depends on the specific repeater scheme). Therefore, many repeater rate simulations assume memory efficiencies of 90\% \cite{sangouard2011}, although some works have considered lower efficiencies \cite{wu2020}. Note that the storage efficiency $\eta_M$ decreases as a function of storage time $\tau$ due to atomic dephasing with some characteristic memory lifetime $T_m$. Assuming a 2-level AFC with Lorentzian-shaped teeth of $T_2$-limited width, one can show that $\eta_M= exp(-\tau/T_M)$ with $T_M=T_2/4$ \cite{jobez2016}.

\emph{Storage time}. The required storage time depends on the repeater protocol and imperfections in the implementation. In the ideal case with large multiplexing, it is given by the time it takes a photon to travel from a repeater node to a link node, and classical information to travel back (see Fig. \ref{fig-repeater} and ref. \onlinecite{sinclair2014}). Assuming an elementary link length of 100 km, a reasonable assumption for a fiber-based quantum network in Europe where cities---and hence access points to the network---are close, this amounts to 500 $\mu s$. However, transmission
loss and device inefficiencies paired with insufficient multiplexing reduce the probability with which entanglement is
heralded across each elementary link within the round-trip time, resulting in the need for longer storage times \cite{wu2020}. For 2-level AFC storage, the storage time is limited by the optical coherence time, see the discussion of the storage efficiency above, while for 3-level AFC it is limited by the spin coherence time \cite{jobez2015}.
    
\emph{Post-selected fidelity}. The fidelity is defined as $F=tr(\bra{\psi}\rho\ket{\psi})$, where $\ket{\psi}$ denotes the input qubit state and $\rho$ the density matrix of the output qubit, conditioned (i.e. post-selected) on the detection of an emitted photonic qubit. The 2-level AFC scheme can be made virtually noiseless---in particular, atomic decoherence does not affect the state of a re-emitted photon \cite{staudt2007,riedmatten2008,horvath2021}---resulting in very high post-selected fidelities, e.g. 99.9\% in ref. \onlinecite{zhou2012}. 
In case of the 3-level protocols (3-level AFC, RASE and NLPE), the storage process is generally not noise free but the fidelity is affected by the quality of the $\pi$ pulses \cite{jobez2015,gundogan2015}. Storage fidelities in the range of 75-85\% have been reported \cite{rakonjac2021,ortu2021} for AFC spin-wave memories and above 90\% for NLPE memories \cite{ma2021b,jin2022}. 
    
\emph{Feed-forward mode-mapping.} This figure of merit addresses the necessity to modify the 
mode of a photon that is entangled with a second photon on the other end of an elementary link after their entanglement has been heralded by the Bell state measurement at the intermediate link node. The goal of the mode-mapping operation is to make the two photons that belong to entangled pairs in neighbouring links indistinguishable so that the subsequent Bell-state measurement, which extends entanglement across these two (or more) links, can be performed (see Fig. \ref{fig-repeater}(c)). In many repeater schemes, temporal multiplexing is assumed and entanglement across neighbouring elementary links is heralded at different times \cite{simon2007,sangouard2011}. In this case, it is necessary to dynamically tune the moment of arrival of the re-emitted photons at the Bell-state measurement. This can be achieved using on-demand readout as in the 3-level AFC protocol, or by means of rapidly switchable optical delay lines \cite{arnold2023}. We emphasize that adjusting the temporal mode is not a general requirement as other degrees of freedom can also be exploited for multiplexing. For instance, ref. \onlinecite{sinclair2014} proposes and partially demonstrates a repeater scheme based on fixed-delay 2-level AFC memories (no temporal mode mapping), spectral multiplexing and frequency shifters plus filters.

\emph{The wavelength of operation.} The wavelengths of rare earth ions that are currently of most interest for quantum memories are $\lambda \approx$ 580 nm (Eu), 606 nm (Pr), 794 nm (Tm), 980 nm (Yb) and 1530 nm (Er). With the exception of erbium \cite{}, they differ from the wavelength at which fiber transmission is maximized. When using a photon pair source based on spontaneous parametric down-conversion or four-wave mixing, this difference can be bridged by designing the source to create non-degenerate photon pairs with one photon optimized for fiber transmission and one for storage\cite{clausen2011,saglamyurek2011}. However, as described in section \ref{subsec:SPDC}, these sources are probabilistic, which impacts the rate with which entanglement can be established across a long and lossy quantum channel. For repeater schemes based on single photon emitters\cite{sangouard2007a} or on frequency-degenerate entangled photons, quantum frequency conversion allows changing the wavelength of the photon travelling to the link node, to the memory, or both, as described in sec. \ref{subsec:sources_figs_merit}.
    
\emph{The storage bandwidth per spectral channel} The storage bandwidth sets the minimum duration of a photon that can be stored. It therefore puts a constraint on the required bandwidth of external photon sources interfaced with the memory. In the case of the AFC protocol, the bandwidth is limited by the energy splitting of the hyperfine states employed for the spectral hole burning, although in some specific cases it can be made larger \cite{etesse2021}. For non-Kramers ions such as Pr and Eu, bandwidths are typically limited by nuclear hyperfine splitting to less than 10 MHz \cite{ortu2022}, but bandwidths in excess of 1 GHz can be achieved in Tm-doped crystals \cite{sinclair2016} due to much larger Zeeman splitting. For Kramers ions such has Yb, the bandwidth can be larger, e.g. 100 MHz in Yb:Y$_2$SiO$_5$ \cite{businger2022}. Finally, we note that by matching the AFC tooth spacing with the level spacing, even larger bandwidths can be obtained, e.g. 5 Ghz in Er:LiNbO$_3$\cite{saglamyurek2011} and 10 GHz in an erbium-doped fiber\cite{wei2024}, however, at the expense of a limited storage efficiency. 
    
\emph{The multiplexing capacity.} For efficient entanglement distribution, the total multiplexing capacity should be large \cite{simon2007,sangouard2011,sinclair2014}. Photons can be stored in a combination of temporal ($N_t$), frequency ($N_f$) and spatial ($N_s$) modes, resulting in a total capacity $N_{tot} = N_t N_f N_s$. For instance, assuming $N_t = 1000$ temporal modes, e.g. for a photon duration of 500 ns and an optical storage time of 500 $\mu$s, $N_f = 100$ frequency modes and $N_s = 10$ spatial modes, we find a total number of modes of $N_{tot} = 1\,000\,000$.

\emph{Reproducibility and ease of use} also play important roles for the creation of large scale networks and for future commercial exploitation. This includes properties like temperature of operation, footprint, and the possibility for integration with other network components. Their solid-state character makes rare earth systems in general very attractive, and the possibility for on-chip integration \cite{saglamyurek2011,rakonjac2022,craiciu2019} yields additional possibilities.

\noindent

\begin{figure*}
\includegraphics[width=0.95\textwidth]{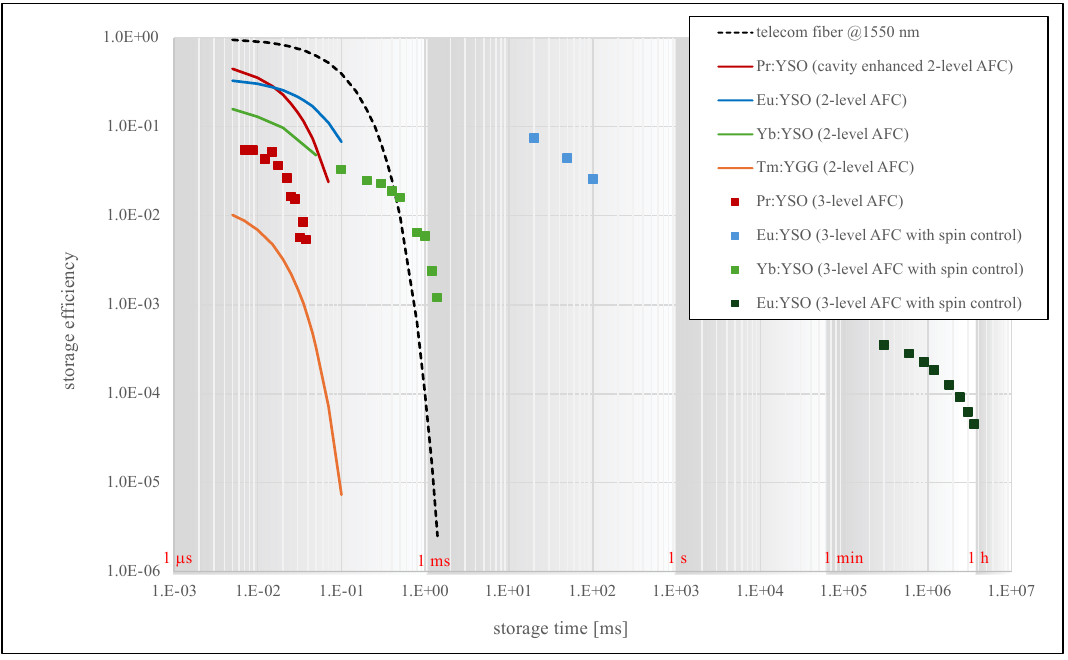}
 \caption{\textbf{Performance of quantum memories.} Storage efficiency (squares: measured data; lines: fits to measured data) versus storage time for different realizations of AFC-based quantum memories in rare-earth-ion doped crystals. The dashed black line additionally shows the case of a standard telecom optical fiber and photons at 1550 nm wavelength, according to eq. \ref{eq:fiber}. The data are based on the following papers: Pr:YSO (cavity enhanced 2-level AFC) -- ref. \onlinecite{duranti2024}; Eu:YSO (2-level AFC) -- ref. \onlinecite{ortu2021}; Yb:YSO (2-level AFC) -- ref. \onlinecite{businger2020}; 
 Tm:YGG (2-level AFC) -- ref. \onlinecite{askarani2021};
 Pr:YSO (3-level AFC) : ref. \onlinecite{rakonjac2021}; 
 Yb:YSO (3-level AFC with spin control) -- ref. \onlinecite{businger2020}. Eu:YSO (3-level AFC with spin control) -- ref. \onlinecite{ortu2021,ma2021}.
Note that we used the standard abreviation YSO for Y$_2$SiO$_5$.}
\label{fig-QMcomparison}
\end{figure*}

 
\subsection{\label{subsec:QM_SOTA} Quantum memory experiments -- state-of-the-art} 

The following section focuses on experimental demonstrations of optical quantum memory using ensembles of rare-earth-ion. It focuses mostly on implementations of the AFC protocol with single or entangled photons, but it also includes notable demonstrations with classical laser pulses and based on other approaches.

Starting with the first experimental demonstration of the 2-level AFC scheme at the single photon level \cite{riedmatten2008}, progress has been very rapid. Key demonstrations of 2-level AFC memories include storage of energy-time \cite{clausen2011,rakonjac2022} or time-bin entangled photons \cite{saglamyurek2011,saglamyurek2015}, storage of single-photon polarization qubits \cite{clausen2012,gundogan2012,zhou2012,jin2015}, storage of hyper entanglement \cite{tiranov2015}, teleportation of a qubit into a memory \cite{bussieres2014,lago2023,iuliano2024}, non-local gates with quantum memory \cite{liu2024}, light-matter entanglement distribution over a metropolitan fiber \cite{rakonjac2023} and entanglement of two AFC quantum memories \cite{usmani2012,grimau2020,lago2021,liu2021}. Temporal multimode storage range from 62-100 stored modes \cite{usmani2010,tang2015,lago2021,ortu2022} to more than 1000 modes \cite{bonarota2011,businger2022,wei2024}. Multimode storage has also been demonstrated using spatial \cite{zhou2015} and spectral \cite{sinclair2014,seri2019} degrees of freedom, as well as a combination of all three degrees-of-freedom \cite{yang2018}. Furthermore, storage efficiencies around 40\%  have been reported \cite{ortu2022,amari2010} with the crystal in a free-space configuration, and cavity-enhanced storage efficiencies of 53-62\% have been achieved for attenuated laser pulses \cite{jobez2014,sabooni2013,duranti2024} and of 27\% for heralded single-photon states \cite{davidson2020}.  The highest reported efficiency of 69 \%  has been demonstrated with GEM \cite{hedges2010}.  

AFC spin-wave storage has first been reported with weak coherent states at the single photon level \cite{jobez2015,gundogan2015} and recently also with entangled photons \cite{rakonjac2021}. A cavity-enhanced AFC spin-wave experiment has reached 12\% efficiency \cite{jobez2014}, and dynamical decoupling has allowed storage of time-bin qubits with a fidelity of 85\% for up to 20 ms at the single-photon level \cite{ortu2021} and storage of classical laser pulses for 1 hour \cite{ma2021}. For the 4-level NLPE memory, spin-wave storage is intrinsic to the protocol, as demonstrated with weak coherent states \cite{ma2021b,jin2022}. Note that RASE \cite{ferguson2016,duda2023} and AFC-DLCZ \cite{laplane2017,kutlu2017,kutlu2019} experiments have demonstrated spin-wave storage of quantum correlations and entanglement.

Erbium-based quantum memories are of particular interest as they operate directly in the telecom C-band at 1532 nm wavelength. Quantum memory protocols have been demonstrated in several Er-doped systems, e.g. with attenuated laser pulses in Er$^{3+}$:Y$_2$SiO$_5$ \cite{lauritzen2010,craiciu2019}, and with entangled photons in an erbium-doped fiber \cite{saglamyurek2015,grimau2020,wei2024}, in Er$^{3+}$:LiNbO$_3$ \cite{askarani2019,zhang2023} and in $^{167}$Er$^{3+}$:Y$_2$SiO$_5$ \cite{jiang2023}. However, Er-doped fiber quantum memories are hampered by short optical coherence times \cite{staudt2006}, while persistent spectral hole burning in Er-doped crystals has so far been limited by spin-lattice relaxation \cite{hastingssimon2008,lauritzen2008}.  

Fig. \ref{fig-QMcomparison} depicts the storage time as a function of the efficiency for several of the above-mentioned quantum memory realizations. While overly simplifying the requirements of a quantum memory for a quantum repeater and focusing only on AFC-based memories, it provides nevertheless some useful insights. The data is extracted from a total of 5 papers, reporting recent results from 2-level AFC-type storage with \cite{duranti2024} and without cavity \cite{askarani2021,businger2020,ortu2021}, and 3-level AFC-based storage with \cite{businger2020,ma2021,ortu2021} and without \cite{rakonjac2021} spin control pulses. We note that we have included demonstrations in which single photons, attenuated laser pulses as well as strong laser pulses were stored. While it is true that storage of strong laser pulses is much easier than storage of true quantum states of light, the protocol steps remain the same. For the purpose of this paper, we therefore ignore this difference.

As a benchmark for comparison, we use the case of a (readily available) telecom optical fiber with absorption coefficient of 0.2 dB/km. The transmission through such a simple delay line---equivalent to the storage efficiency of a quantum memory with fixed delay--- is given by 

\beq
t_{fiber}=10^{-0.02\,\ell}
\label{eq:fiber}
\eeq
where $\ell$ is the fiber length in km. Clearly, for an atomic quantum memory to be useful, its efficiency has to be higher than the fiber transmission for the same delay. Several observations are noteworthy.

For storage times below around 400 $\mu$s, no memory implementation currently beats the simple fiber delay line in terms of storage efficiency. However, the gap has narrowed significantly over the past few years, e.g. to less than a factor of two in the case of cavity-enhanced memory with small storage time \cite{duranti2024}. For storage times beyond 400 $\mu$s, the atomic memories perform better than the fiber delay line. Still, the storage efficiencies need to be improved to make these memories useful for quantum repeaters.

The efficiencies in the implementations of the 2-level AFC protocol in materials with optical coherence times above 1 ms \cite{askarani2021,bussieres2014,ortu2021} decrease faster than what one would expect from this coherence time. This currently limits storage times to around 100 $\mu$s even though memory lifetimes $T_M$ around 300 $\mu$s should in principle be achievable. The reasons for the reduced lifetime are spectral diffusion, cryostat vibrations, and laser line jitter. These issues also limit the time during which quantum information can be stored in optical coherence in the 3-level AFC protocol \cite{ortu2022}, which in turn limits the multimode capacity.

Assuming that the first quantum repeater generation will employ probabilistic SPDC-based entangled photon pairs with small photon pair generation probability, the required memory storage times will be long compared to the round-trip time from a repeater node to a link node and back \cite{simon2007}. This is required to ensure that heralded entanglement across adjacent elementary links will eventually be created. Such storage times---around 100 ms in ref. \onlinecite{wu2020}---will require using the 3-level AFC protocol with spin wave control such as in ref. \onlinecite{ortu2021}, as well as cavity-enhanced light-matter interaction as in ref. \onlinecite{duranti2024}.

Once efficient and sufficiently multimode photon sources are available, shorter storage times will suffice. Ideally, the storage time equals the round-trip time mentioned above. 
\begin{itemize}
    \item For fiber-based quantum communication in densely-populated areas, it makes sense to space link nodes at intercity distances, e.g. between a few tens of kilometres to hundred kilometres. This yields storage times of a few hundred $\mu$s, which is possible using the 3-level AFC protocol but is also getting into reach of 2-level AFC implementations.
    \item For quantum communication through long-haul fibers or operating over satellite-based communication links, the distance between nodes and hence the round-trip times will be much longer, e.g. a few thousand km and around 10 ms in the case of a satellite, respectively. These storage times require the use of the 3-level AFC protocol together with spin-wave control. First demonstrations\cite{ortu2021, ma2021} show that this may be feasible in near future. 
    \item Finally, one could also assume that the memory itself is transported to exchange quantum information. While the transportation speed depends strongly on the vehicle---a car, a plane, satellites with different orbital speeds---it is clear that storage times of at least 1 h will be required. Surprisingly, the work reported in ref. \onlinecite{ma2021} shows that this may be feasible, but storage efficiencies still require significant improvements.  
\end{itemize}

In summary, we find that quantum memories based on ensembles of rare-earth ions are likely to become rapidly useful for quantum repeaters over optical fibers and free-space links, and hence also for extended quantum networks based on a combination of those. 


\section{\label{sec:nodes}Network components : repeater nodes with quantum processing capability. }

Previously we described repeater nodes based on ensemble memories. In this section, we describe repeater nodes composed of individual ions that encode long-lived qubits and allow quantum information processing. 

The potential of rare-earth ions for quantum gates and quantum computing was first pointed out in 2002 \cite{ohlson2002} (see ref. \onlinecite{kinos2021} for a recent review). Qubits can be encoded into nuclear spin states of individual ions, whose coherence times in strong magnetic fields can reach 6 hours \cite{zhong2015}. Furthermore, ion spacing of as little as a few nanometers---either naturally or through deterministic ion implantation \cite{groot2019}---enables efficient qubit-qubit interactions. The proximity, however, creates a problem with the need for individual addressing. This can be circumvented in an elegant manner by using optical control pulses, which results in spectral selectivity due to the inhomogeneous broadening of the optical transitions \cite{ohlson2002}. Infidelities for single and controlled two-qubit gates have been predicted to be of around $10^{-4}$ and $10^{-3}$, respectively \cite{kinos2021b, kinos2022}, allowing one in principle to create high-fidelity spin-photon entanglement and in turn entanglement between distant spins using the Barrett-Kok scheme \cite{barrett2005}. In addition, using controlled interactions between neighbouring ions, it may be possible to reach the noisy intermediate scale quantum (NISQ) domain \cite{preskill2018}. Initial experimental work focused on demonstrations using large ensembles \cite{walther2009}, but this does not scale to multiple qubits \cite{wesenberg2007}. Instead, scalable quantum information processing requires the individual ion regime, which has historically been difficult to reach with rare-earth systems due to their long excited-state lifetimes. However, in recent years there has been significant progress, and many research groups are now routinely investigating single rare-earth ions. Towards this end, it is important to collect a strong fluorescence signal, which can be achieved using cavity-enhanced transitions and the Purcell effect discussed in Sec. \ref{sec:Purcell}.

The recent breakthroughs in detecting single ions allow for a different type of repeater nodes than what has been discussed above, more precisely repeater nodes in which quantum information is stored in qubits capable of information processing. We will refer to these nodes also as quantum processor nodes. In view of building a quantum repeater, the main advantage of using quantum processor nodes is that they open up the possibility for deterministic---as opposed to probabilistic\cite{lutkenhaus1999}---entanglement swapping operations between neighbouring elementary links, however, at the expense of less multiplexing across each individual elementary link.
 
Quantum repeaters based on individual ions employ three main steps, which are illustrated in Fig. \ref{fig-repeater2}, briefly described below, and discussed further in Secs. \ref{subsec:QPE} to \ref{subsec:DQQEO}. 

\begin{itemize}
    \item \emph{Qubit-photon entanglement}: Individual qubits in repeater nodes are entangled with single photons. 
    \item \emph{Photonic entanglement swapping}: Through photon interactions and measurements, the qubit-photon entanglements are swapped into entanglement between qubits in adjacent nodes.
    \item \emph{Deterministic qubit-qubit entanglement swapping}: Joint operations and measurements on two neighbouring qubits in the same quantum repeater node---each entangled with a qubit in another node---enable deterministic entanglement swapping across adjacent elementary links.
\end{itemize}


\subsection{\label{subsec:QPE}Qubit-photon entanglement}
This section discusses two main ways in which qubits can be entangled with single photons. In the first method, qubits are put into superposition states and used as emitters of single photons so that the state of the ion is entangled with the emitted photon. This protocol may rely on an individual ion acting as both the photon emitter and the long-lived qubit, as in the case of trapped ions \cite{sangouard2009} and diamond color centers \cite{hensen2015}. Ideally, the emitter features a Purcell-enhanced optical transition that should additionally be at telecom wavelength, although it is possible to frequency convert photons at the cost of added complexity (see section \ref{subsec:sources_figs_merit}. However, Purcell enhancement reduces the excited state lifetime, while the controlled two-qubit gate assumed in ref. \onlinecite{kinos2023a} and described in Sec. \ref{subsec:DQQEO} benefits from long lifetimes and narrow optical transitions. Thus, it can be beneficial to use two different ion species: one for communication and one for storage \cite{asadi2018, debnath2021, kinos2023a}. Attractive candidates for communication ions are Er due to its telecom transition and Nd due to its large oscillator strength in many hosts \cite{alqedra2023}, whereas Eu is a strong candidate for encoding qubits due to its long spin lifetimes \cite{koenz2003} and spin coherence times \cite{zhong2015}. Other potentially interesting ions are Yb due to its reduced sensitivity to magnetic fields---Yb features a zero first-order Zeeman (ZEFOZ) transition at zero magnetic field \cite{ortu2018}---and Tm, which, in certain crystals, features zero-phonon lines that connect the ground state with different excited states \cite{alfasi2022}, one of which could be Purcell enhanced and one of which could be kept spectrally narrow.  

In the second method, a single photon impinges on a cavity whose properties depend on the state of a qubit located inside of it. This can either be a change of reflection vs transmission of the photon \cite{knaut2024} or a change of the phase of the reflected photon \cite{wade2016, welte2018}. Note that these demonstrations, which were done using diamond color centers or trapped atoms, can be generalized to rare-earth ions.


\subsection{\label{subsec:PPES}Photonic entanglement swapping}
Consider the case when qubit-photon entanglement is created using the first method described in the previous section, i.e., where qubits are used as emitters of single photons. Qubits in adjacent repeater nodes can then be entangled by sending the photons to a link node where they are subjected to a probabilistic linear-optics Bell-state measurement, as in the ensemble-based scheme discussed above. This swaps the entanglement from two qubit-photon pairs in a heralded fashion to the qubit-qubit pair. Without additional resources this process is probabilistic, however, please recall that this measurement---but not the transmission of photons across the elementary link---can in principle be achieved deterministically by adding auxiliary entangled photons \cite{grice2011,barz2023}. 

When both communication and storage ions are used, the end goal is to entangle storage ions, and we will now describe two approaches to achieve this. Either one first entangles two communication ions in adjacent repeater nodes using, for instance, the Barrett-Kok scheme \cite{barrett2005,hensen2015}, and then one swaps the entanglement to the storage ions \cite{asadi2018}; or one starts by locally entangling the storage ion with the photon emitted from the communication ion and then attempts entanglement swapping by measuring two photons in a link node \cite{kinos2023a}. In the former case, the communication ions must remain idle with good coherence properties during the time it takes the photons to travel to the link node and the classical heralding information to return. Provided that entanglement has been established across the elementary link, it is swapped to the storage ions, and one can attempt to entangle the communication ion with another processor node. The benefit of this approach is that it only requires one communication ion and one storage ion per node. Conversely, the latter method has the advantage of freeing up the communication ion immediately after it has emitted a photon, thus relaxing the coherence requirements of the communication ion. Furthermore, if additional storage ions are available, the protocol allows for time multiplexing by reusing the communication ion to emit new photons entangled with other storage ions. 

Alternatively, if the second method described in the previous section is used, qubit-qubit entanglement can be achieved via a deterministic entangling operation implemented by means of state-dependent reflections of a time-bin photonic qubit from a cavity coupled to an individual ion \cite{knaut2024}.


\subsection{\label{subsec:DQQEO}Deterministic qubit-qubit entanglement swapping}
Deterministic entanglement swapping between qubits in repeater nodes can be achieved as long as two-qubit gates and qubit measurements can be performed on the two qubits involved in the entanglement swapping in a node \cite{gottesman1999}. 

\begin{figure}
\includegraphics[width=0.8\columnwidth]{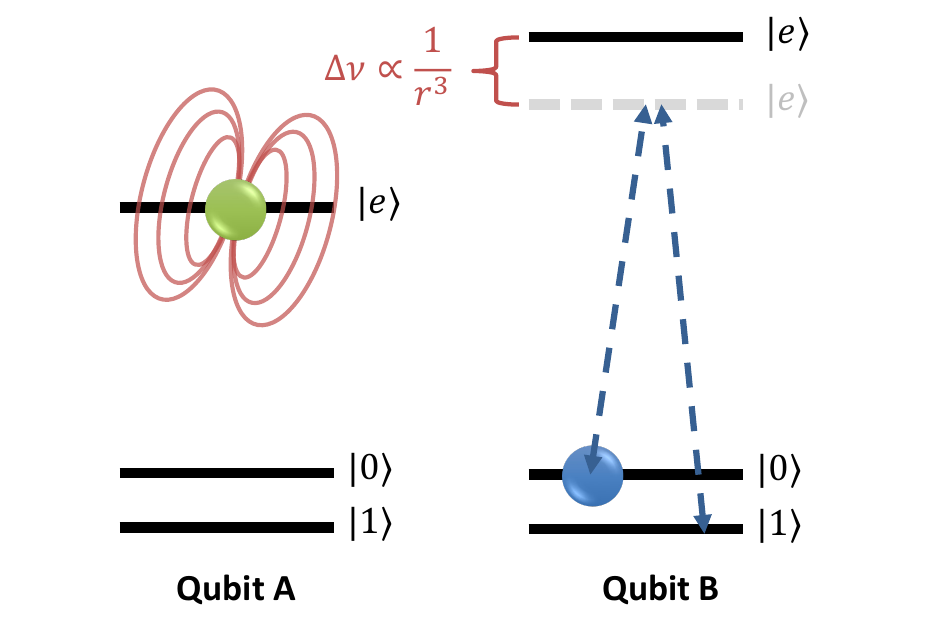}
\caption{\textbf{Electric dipole blockade protocol.} The protocol exploits that an excitation of qubit A (the control qubit) changes its static electric dipole moment, thus altering the electric field surrounding the qubit. In turn, this causes a frequency shift of the transition frequencies of nearby qubits (the target qubits), scaling as $1/r^3$, where $r$ is the distance between the qubits. If this shift is sufficiently large, any laser pulses originally resonant with the transitions of the target qubit would no longer affect the qubit. This blocks the operation. }
\label{fig-blockade}
\end{figure}

Provided the two qubits are sufficiently close and that their states feature permanent electric dipole moments, two-qubit gates can be implemented using the electric dipole-blockade, which is explained in Fig. \ref{fig-blockade}. For example, a CNOT gate can be realized by first exciting the control qubit from $\ket{0}\rightarrow\ket{e}$, then attempting a NOT operation on the target, before finally de-exciting the control from $\ket{e}\rightarrow\ket{0}$. Due to the blockade effect, the NOT operation on the target is only possible if the control is in state $\ket{1}$. The key to this gate is that the transition of the target qubit from $\ket{0}$ to $\ket{e}$ is spectrally sufficiently narrow to ensure that a small Stark shift $\Delta\nu$ detunes it out of resonance with the lasers used for the NOT operation. Alternatively, magnetic dipole-dipole interactions can be used, either between weakly interacting spins of two adjacent rare-earth ions or between the electron spin qubit and nearby nuclear spins of the host. This basic interaction has recently been demonstrated in rare-earth systems \cite{uysal2023, ruskuc2022}. 

The capability of performing two-qubit gates additionally allows for entanglement purification protocols \cite{deutsch1996, dur1999, bratzik2013, krastanov2019}, and if there are sufficiently many storage qubits and the operations are sufficiently good, one can even consider performing error correction within nodes \cite{jiang2009, munro2010} to improve the fidelity and efficiency of the quantum network. Initial theoretical work that focuses on the specific properties of rare-earth ions has already been reported \cite{rolander2022}.

\subsection{\label{subsec:QNNFoM}Figures of merit}
This section focuses on the figures of merits of quantum processor nodes. Some of these merits are similar to the ones already discussed for quantum memories---including the wavelength of operation and the storage time---whereas others are different, and those are briefly discussed here. Importantly, in contrast to the case of quantum memories, the storage efficiency is no longer relevant, and atomic decoherence impacts the fidelity instead.  

\emph{Fidelity of the entangled operation between two nodes.} It is possible to use SPDC sources for qubit-based protocols, for example using the photon reflection techniques \cite{knaut2024} with cavities. In this case, there is a trade-off between the entanglement fidelity and the entanglement rate, as discussed in Sec. \ref{subsec:SPDC}. For repeaters based on qubits that emit single photons, this trade-off is removed, and the entanglement fidelity is instead limited by the quality of the single photon emission---single photon purity and indistinguishability between photons---together with the fidelity of the qubit gate operations.

\emph{Multiplexing capacity.} Multiplexing can be achieved by duplicating the setup in each quantum processor node, i.e. by adding more cavities---each with its own communication ion and potentially storage ions---that can be used in parallel. As discussed earlier, time multiplexing can also be achieved if single photons emitted by the communication ion are directly entangled with storage ions. 

\emph{Photon bandwidth.} In protocols that use only a single ion type as both emitter and qubit, the photon bandwidth may be limited by the energy level spacing between the qubit levels, in similarity with some of the ensemble-based protocols. Protocols that directly entangle emitted photons with storage ions avoid this limitation. However, a similar restriction arises if the communication ion interacts with the qubit ion using the dipole blockade mechanism, as the dipole frequency shifts must be larger than the bandwidth of the optically excited state of the communication ion.


\subsection{\label{subsec:QNN_SOA}Quantum processing nodes -- state-of-the-art}
Despite the first demonstration of cavity-enabled single ion detection being only a few years old, there have already been significant advances. Several groups have achieved single-shot readout of individual spin qubits, for example in Er-doped silicon \cite{gritsch2024}, Er-doped Y$_2$SiO$_5$ \cite{raha2020}, Yb-doped YVO$_4$ \cite{kindem2020}, and four Er ions doped in Y$_2$SiO$_5$ have also been simultaneously initialized and read out in ref. \onlinecite{chen2020}. Furthermore, coherent spin manipulation of individual ions has also been demonstrated \cite{chen2020, kindem2020}. 

In ref. \onlinecite{ruskuc2022}, a Yb ion was coupled to an ensemble consisting of four equidistantly spaced V host ions, and via a spin-exchange interaction, the authors could generate collective spin excitations of the ensemble and use it as a quantum memory. Furthermore, they were able to prepare and measure maximally entangled Yb-V Bell states. 

Controlled interactions have also been shown in Er-doped Y$_2$SiO$_5$ \cite{uysal2023}, where a single Er ion interacted with a $I=1/2$ nuclear spin in the host crystal, identified as a fortuitously located proton ($^1$H). Through the use of dynamical-decoupling sequences applied to the electron spin, the nuclear spin could be controlled and both single- and two-qubit gates performed. 

Another crucial aspect required for repeater nodes based on quantum processors is the interaction between individual ions and single photons. As discussed previously, this can be achieved through state-dependent reflections of photons from a cavity coupled to an individual ion \cite{knaut2024}, but a demonstration of such a protocol using rare-earth ions is still missing. However, entanglement between an individual Er and an emitted photonic time-bin qubit has recently been reported \cite{uysal2024}. Furthermore, two Yb ions located in two separate nanophotonic cavities have been entangled via a joint measurement of two emitted photons \cite{ruskuc2024}. In this latter work, the authors also demonstrated probabilistic quantum state teleportation between the two Yb qubits, and generated a tripartite W-state between three Yb spin qubits (two located in one nanophotonic cavity and one located in a second cavity).


\section{\label{sec:elementary-repeaters}Elementary quantum repeater links}

In this section, we will briefly discuss experiments with rare-earth systems that aim at the demonstration of an elementary quantum repeater link, i.e. experiments where at least two quantum memories were entangled. Note that demonstrations of entanglement between photons and a single quantum memory are already discussed in Sec. \ref{subsec:QM_SOTA}.

Entanglement between two separate rare-earth systems was reported for the first time in 2012 \cite{usmani2012}. A single-photon entangled state was generated by splitting a heralded photon, generated by means of spontaneous parametric down-conversion (SPDC), between two spatial modes. Each mode was stored in a separate quantum memory using the 2-level AFC protocol, resulting in a single excitation delocalized---or entangled---between two Nd$^{3+}$:Y$_2$SiO$_5$ crystals separated by 1.3 cm. The storage time in this demonstration was 33 ns. While this creation of memory-memory entanglement was heralded by a photon detection, the scheme is not scalable to large distances between the crystals without loosing the efficiency of the heralding process.

In 2020, another group reported the storage of both entangled photons from an SPDC-based source, again employing the 2-level AFC protocol \cite{grimau2020}. In this demonstration the two photons featured different wavelengths, requiring the use of two different rare-earth memories : a Tm-doped crystal for the 794 nm photons and an Er-doped crystal for the 1535 nm photons. Storage times in this proof-of-principle demonstration---32 ns and 6 ns, respectively---were again short. In this case, the storage process was not heralded, as opposed to the requirement for an elementary repeater link described in section \ref{subsec:repeaters}.

Several shortcomings of these two demonstrations were overcome in 2021, when two groups demonstrated the heralded creation of entanglement between two crystals, in line with the repeater architecture depicted in fig. \ref{fig-repeater} \cite{lago2021, liu2021}. In both demonstrations, each of two SPDC sources emitted one photon per pair into a 2-level AFC quantum memory and the second photon over optical fiber to a central linear optics Bell-state measurement. Subsequent photon detection then heralded entanglement between the two memories.

In ref. \cite{liu2021}, two Nd$^{3+}$:VO$_4$ quantum memories were employed to store photons at 880 nm during 56 ns, and the Bell-state measurement was based on the detection of 2 photons. While the demonstration did herald entanglement between the stored and subsequently recalled photons, the combination of such a two-photon Bell-state measurement with photon pairs generated by means of SPDC implies that this elementary link is not scalable \cite{guha2015}.

In ref. \cite{lago2021}, a single-photon Bell-state measurement was employed instead, which allows scalable concatenation of elementary links. Single-photon schemes, on the other hand, complicate the entanglement verification and require interferometric stability of the fiber link \cite{chou2005}. The demonstration relied on two Pr$^{3+}$:Y$_2$SiO$_5$ memories that stored photons at 606 nm wavelength up to 25 $\mu$s, including in multimode fashion. Note that the wavelength of the second member of each photon pair was 1436 nm, which, in principle, would allow long-distance entanglement.  

Even more recently, the first demonstration of an elementary repeater link based on individual rare-earth ions was reported \cite{ruskuc2024}. This work is particularly noteworthy as it showed for the first time that spectral diffusion can be overcome through the combination of a crystal with rare-earth sites that feature reduced sensitivity to electric field noise (see discussion in sec. \ref{subsec:sources_SOTA}) and active spectral control. Using two Yb ions in separate YVO$_4$ crystals and a Bell-state measurement based on the detection of a single photon, the researchers did not only demonstrate heralded entanglement that lasted for almost 10 ms in a scalable fashion, but also quantum teleportation as well as the creation of a tripartite entangled state, which required adding a third ion.

To finish this brief review of developmets related to the creation of elementary quantum repeater links, let us also mention the demonstration of interference of weak laser pulses recalled from separate Tm:LiNbO$_3$-based quantum memories \cite{Jin2013}. In contrast to the above-described experiment,  this demonstration targeted the connection between neighbouring elementary repeater links.


\section{\label{sec:outlook}Conclusion and Outlook}
In this review we have detailed the progress towards quantum repeater networks made so far using the rare-earth-ion based platform. Here we briefly conclude by providing some suggestions for future work.

\emph{Single emitter:} Starting with the first demonstrations of Purcell-enhanced emission in 2018, the development of single rare-earth ions as sources of individual photons has been very rapid, and emitters at telecom wavelength around 1532 nm (Er), at around 980 nm (Yb), and at around 880 nm (Nd) have been demonstrated. However, some of the best rare-earth-ion based quantum memories employ Pr, Eu and Tm, and more work is required to develop photon emitters using these ions. This will allow overcoming the limitations of SPDC-based sources in an approach that ensures spectral compatibility with ensemble-based quantum memories. In addition, and regardless of the particular emitter chosen, it is important to better understand, and subsequently avoid or control spectral diffusion. This will enable multi-photon interference  and in turn allow creating heralded entangled photons using four single photons as well as entanglement of qubits---also encoded into singe rare-earth ions---over large distances.

\emph{Ensemble memories:} Owing to their large multimode storage capacity, long storage times, high fidelity and high storage efficiency, ensemble-based rare-earth memories have established themselves during the past 16 years as key systems for quantum repeaters. An important challenge for continued progress is to achieve all key requirements for quantum networks and repeaters with a single device. For instance, an ambitious yet reasonable goal would be a quantum memory with $\geq$50\% efficiency, $\geq$ 10$\,$000 mode capacity, and $\geq$ 100 $\mu$s optical storage time with the possibility of extending the storage time to 100 ms through spin-wave storage. 
Other challenges include optimizing the entire memory system including cryostat and laser source to enable future field deployment. Towards this end, an interesting possibility is the creation of integrated (on-chip) devices following pioneering work in Tm:LiNbO$3$\cite{saglamyurek2011} but exploiting the new waveguide fabrication possibilities that underpin the creation of nano-photonic cavities.

\emph{Individual qubits:} Over the past few years, the number of investigations related to interactions between photons and qubits encoded into individual rare-earth ions has increased rapidly, and qubit read-out as well as early stages of control and gate operations have been reported. This demonstrates that nano- and micro-cavity technology has reached the point where the limitations due to the long excited state lifetimes can be overcome and the benefits of these states for gate operations can finally be exploited. This work benefits from similar demonstrations with other emitters, including diamond vacancy centers and trapped atoms, that can be generalized conveniently to the new rare-earth systems. 
While interactions between rare-earth ions and surrounding lattice spins have been realized in a few groups, coupling between two (or more) rare-earth ions that enables controlled multi-qubit gates still remains to be demonstrated -- an important goal for the near future. When such operations are available, it will become possible to progress to small processor nodes with a dense set of spectrally distinguishable qubits, in similarity with what has been achieved with spins in NV centers. But in comparison, a rare-earth processor node would offer better optical coherence as well as stronger ion-ion interactions, resulting in higher operation bandwidths and many more spectral channels.

In this paper, we have reviewed properties of rare-earth-ion doped crystals that make them appealing and, due to their long optical coherence time, arguably unique candidates for light-matter interfaces needed in future quantum networks and especially in quantum repeaters. Furthermore, we have discussed recent progress towards the creation of three different network components -- single photon emitters and long-lived qubits based on individual ions, as well as multi-mode quantum memory for light based on large ensembles of ions. The fact that all of these can be based on the same material system suggests that it is possible to combine several components into a single, robust quantum photonic integrated chip (QPIC).
But in addition to pushing technological maturity, we stress that continued basic material research and spectroscopic studies are paramount to the development of high-performance quantum networks. Indeed, the recent years have shown that exploring new operational regimes---different magnetic fields, doping concentrations etc.---or new combinations of rare-earth ions and host crystals sometimes leads to better properties such as reduced sensitivity to environmental perturbations. The rich diversity of rare-earth systems is unique in this context, and we are convinced that we will witness more, and more-advanced poof-of-principle demonstrations of quantum network technology, possibly including the first scalable quantum repeater, in the near future.\\


\textbf{Acknowledgements.} This work has received funding from the Swiss State Secretariat for Education, Research and Innovation (SERI) under contract number UeM019-3 (W.T. and M.A.) and contract number UeM029-7 (MA). A.K. acknowledges support from the Wallenberg Centre for Quantum Technology (WACQT) funded by Knut and Alice Wallenberg Foundation (KAW). A.W. ackowledges support from the Swedish research council (grant no. 2021-03755) and the Olle Engkvist Foundation.

\bibliography{RE-review}

\end{document}